\documentclass[sigconf]{acmart}
\usepackage{hyperref}

\usepackage[normalem]{ulem}
\usepackage{graphicx}
\usepackage{tgcursor}  
\usepackage{epigraph}
\usepackage{tikz}
\usepackage{xcolor}
\usepackage{amsfonts,amsmath,amsthm}
\usepackage[mathcal]{eucal}
\usepackage[shortlabels]{enumitem}
\usepackage{tablefootnote}
\usepackage{caption}
\usepackage{filecontents}
\usepackage{lipsum,afterpage,refcount}

\AtBeginDocument{%
  \providecommand\BibTeX{{%
    \normalfont B\kern-0.5em{\scshape i\kern-0.25em b}\kern-0.8em\TeX}}}

\setcopyright{acmcopyright}
\copyrightyear{2021}
\acmYear{2021}
\acmDOI{}

\acmConference[Seoul'21]{Seoul'21: ACM Conference on Computer and Communications Security (CCS)}{November 14--19, 2021}{Seoul, South Korea}
\acmBooktitle{Seoul'21: ACM Conference on Computer and Communications Security (CCS), November 14--19, 2021}
\acmPrice{}
\acmISBN{}

\begin{document}

\title{Ontology-driven Knowledge Graph for Android Malware}



\author{Ryan Christian$^1$, Sharmishtha Dutta$^1$, Youngja Park$^2$, Nidhi Rastogi$^1$}
\affiliation{%
  \institution{Rensselaer Polytechnic Institute$^1$, IBM TJ Watson Research Center$^2$}
  \country{New York, USA}
}
\renewcommand{\shortauthors}{Anonymous}

\begin{abstract}

We present MalONT2.0 -- an ontology for malware threat intelligence \cite{rastogi2020malont}. New classes (attack patterns, infrastructural resources to enable attacks, malware analysis to incorporate static analysis, and dynamic analysis of binaries) and relations have been added following a broadened scope of core competency questions. MalONT2.0 allows researchers to extensively capture all requisite classes and relations that gather semantic and syntactic characteristics of an android malware attack. This ontology forms the basis for the malware threat intelligence knowledge graph, MalKG, which we exemplify using three different, non-overlapping demonstrations. Malware features have been extracted from CTI reports on android threat intelligence shared on the Internet and written in the form of unstructured text. Some of these sources are blogs, threat intelligence reports, tweets, and news articles. The smallest unit of information that captures malware features is written as triples comprising head and tail entities, each connected with a relation. In the poster and demonstration, we discuss MalONT2.0, MalKG, as well as the dynamically growing knowledge graph, TINKER.

\end{abstract}

\ccsdesc[500]{Computer systems organization~Embedded systems}
\ccsdesc[300]{Computer systems organization~Redundancy}
\ccsdesc{Computer systems organization~Robotics}
\ccsdesc[100]{Networks~Network reliability}

\keywords{Knowledge Graphs, Security Intelligence, Android, Malware}


\maketitle

\section{Introduction}
Malware attack intelligence describes the working of the attacks, their tactics, techniques, and procedures (TTPs), and the technology vulnerabilities exploited by the malware. This intelligence can equip security researchers with information to build better defenses against advanced cyber attacks and issue early warnings about future threats. The Internet can lead to evidence on attack intelligence through thousands of diverse and heterogeneous sources, globally known as cyber threat intelligence (CTI). Discerning and utilizing this knowledge speedily and accurately for longitudinal studies mandate rigorous development of techniques that are constantly evolving and adapting to the complexities of malware attacks. Therefore, threat intelligence is best utilized when it is timely, actionable, shared in an universally acceptable format, contextual, and understandable.
\par
CVE\footnote{https://cve.mitre.org/}, NVD\footnote{https://nvd.nist.gov/} are vulnerability tracking programs where information is combined through a centralized platform in a semi-structured way. Industry standards like Structured Threat Information eXpression (STIX)\cite{stix} and trusted automated exchange of indicator information (TAXII)\cite{connolly2014trusted} provide a language-agnostic framework for storing, sending, and receiving packages. However, despite concerted efforts by experts towards organizing malware threat information, there is a lack of context and transparency in sharing CTI. Analysts require more than data-driven threat intelligence. They seek trustworthy data sources, relevant threat indicators, sources and motivations behind attacks, and the likelihood of an attack.  They also demand context to get a comprehensive picture of the threats, victims, and the distinct tactics an attacker deploys.
\par
Malware threat ontologies, like MalONT2.0, enable communicating contextual CTI feeds by representing them in a structured format that encapsulates data and information characterized by properties that vary according to context. Our main contributions are as follows:
\begin{enumerate}
    
    \item We present MalONT2.0, an ontology for capturing malware threat intelligence through classes and relations that combine semantic features (such as malware, attacker, infrastructure) with syntactic features and factual data (extracted from VirusTotal\footnote{virustotal.com}).
    
    \item Annotated CTI reports on android malware attacks, where each annotation is instantiated into classes and shares a relationship with other instances. Both classes and relations are defined and described in MalONT2.0 and share similarities with the STIX2.1 framework, where applicable. The instances are stored as RDF triples \cite{pujara2013knowledge}, $\langle entity_{head}, relation, entity_{tail}\rangle$.
    
    \item We provide a dynamically growing knowledge graph generated by automatically feeding new CTI reports into the existing KG. We demonstrate the use of this knowledge graph through three queries (shown in Section \ref{demo}).

\end{enumerate}

\subsection{What's new in MalONT2.0}\label{newMalONT}
MalONT2.0 is a significant improvement over prior version \cite{rastogi2020malont} as it comprehensively captures semantic, syntactic, and factual description of a malware threat. Prior version emphasised on contextual information coming from semantic data and contained partial factual data. The main source of CTI was unstructured threat reports written on diverse threats such as malware, and APT. The latest knowledge graph is generated from CTI reports that focus exclusively on android malware threats. Designing an ontology requires answering competency questions that can provide a wide coverage of the domain. These questions (or a narrower version of a competency question) are validated by confirming answers to queries on the instantiated triples. While building MalONT2.0, we updated the competency questions to the following:

\begin{enumerate}[a]
    
    \item Find all missing intelligence from the KG relating to various attack vectors - malware, actors (attacker, attacker-group, organizations, country), infrastructure (software, applications, platform, infrastructure used, TTPs). This information may be spread across triples generated across multiple CTI reports describing an attack vector.\label{CQ1}
    
    \item  Triples collected from CTI reports from a wide range of sources can be aggregated with syntactic intelligence from a source like VirusTotal to provide a richer description of the attack vector.\label{CQ2}
    
    \item Identifying similar properties and grouping attack vectors can reveal latent behaviors and can be used in predictive models to forecast future events, both short- and long-term.\label{CQ3}

\end{enumerate}
        

\begin{figure}[hbt!]
        \centering

        \frame{\includegraphics[width=0.2\textwidth, height=0.2\textwidth]{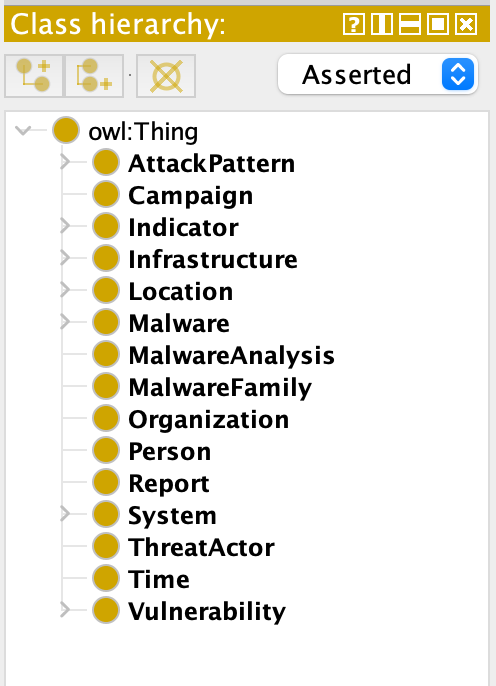}}
        \hspace{.5cm}
        \frame{\includegraphics[width=0.2\textwidth, height=0.2\textwidth]{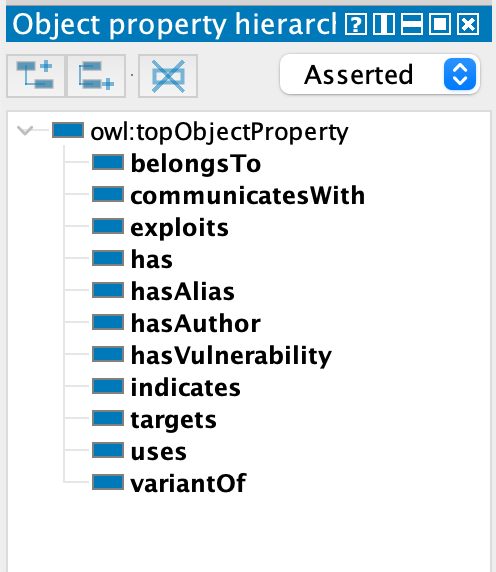}}
        \hspace{.5cm}
        \caption{Main Classes (left), Relations (right) from MalONT2.0}
        \label{fig:ontology}
\end{figure}

\subsection{Example Knowledge Graph}
See Figure~\ref{fig:snap_kg_2} (left) for a sub-graph of MalKG. A CTI report contains information about a malware, Pegasus mapped to \textsf{Malware} class. This malware is also known as Chrysaor, and it logs user keystrokes and leaks the data of popular apps. The CTI report contains the hash for a single sample of the malware. According to VirusTotal, this sample was first seen in April 2017.

\section{System Architecture}
The proposed framework dynamically gathers unstructured CTI reports from the Internet. It extracts threat intelligence information in the form of RDF triples, assigns them classes and relations from MalONT2.0 (see Figure~\ref{fig:ontology}) forming a new knowledge graph, which it appends to the existing MalKG. In this section, we describe the main components of this framework .

\begin{enumerate}
     
\item\textbf{CTI reports corpus}-- MalONT2.0 is used to instantiate 25 CTI reports written between 2011 -- 2021, and downloaded from the Internet. We followed the process of natural language annotation for machine learning \cite{pustejovsky2013natural}, created mutually agreed upon annotation guidelines, including a tie-breaking process managed by a security expert. These reports were authored by analysts from security organizations such as McAfee Labs. The annotated text has approximately 3,400 tags extracted by annotators using BRAT\footnote{https://brat.nlplab.org/} resulting in 1,100 entities and 2,300 relations. Triples generated from these  are the structural components of MalKG that capture large-scale facts related to android malware threat intelligence.
\item\textbf{MalONT2.0}-- Semantic text patterns map to classes (called entity in KG) and object properties (called relations in KG) defined by MalONT2.0. An ideal open-source ontology can systematically capture cyber threat and attack information (facts and analysis) to model the contents of CTI reports. For instance, in MalONT2.0, three classes largely describe malware behavior -- \textsf{Malware}, \textsf{Vulnerability}, and \textsf{Indicator}. Instances of these classes can equip the analysts with information on malware behavior and TTPs. See Figure~\ref{fig:ReportTriples} for a snapshot on an instance of MalONT2.0. See GitHub\footnote{https://github.com/aiforsec/DemoCCS2021/} code for annotated text and corresponding triples for all CTI reports.

\item\textbf{Knowledge Graph Generation and Querying}--
We construct a knowledge graph corresponding to the malware ontology by populating it with triple instances derived from actual CTI reports. However, one may argue about the necessity concerning knowledge graphs, given that the instantiated ontology is previously obtaining facts and knowledge regarding the domain. The two key features of knowledge graphs are the ability to reason on deduced information and infer latent information. MalKG captures properties connecting nodes (also called entities) and employs a reasoner to draw associations among entities that would otherwise not be recognized.

\item\textbf{Dynamically generated KG}-- In addition to the MalKG generated from annotated triples, we also present a dynamically growing knowledge graph called TINKER. CTI reports are regularly published by security and technology companies. These reports, especially those freely available for public access, are shared on the Twitter platform by company personnel. We have built an interface using python that uses academic Twitter API to extract unique occurrences of android malware CTI reports. Triple extraction models trained on annotated instances are used for dynamic extraction from frequent batches of CTI reports.
\end{enumerate}

\section{Outline of Poster and Demonstration} \label{demo}
\begin{figure}[hbt!]
    \centering
 \frame{
   \includegraphics[width=.4\textwidth,height=.25\textwidth]{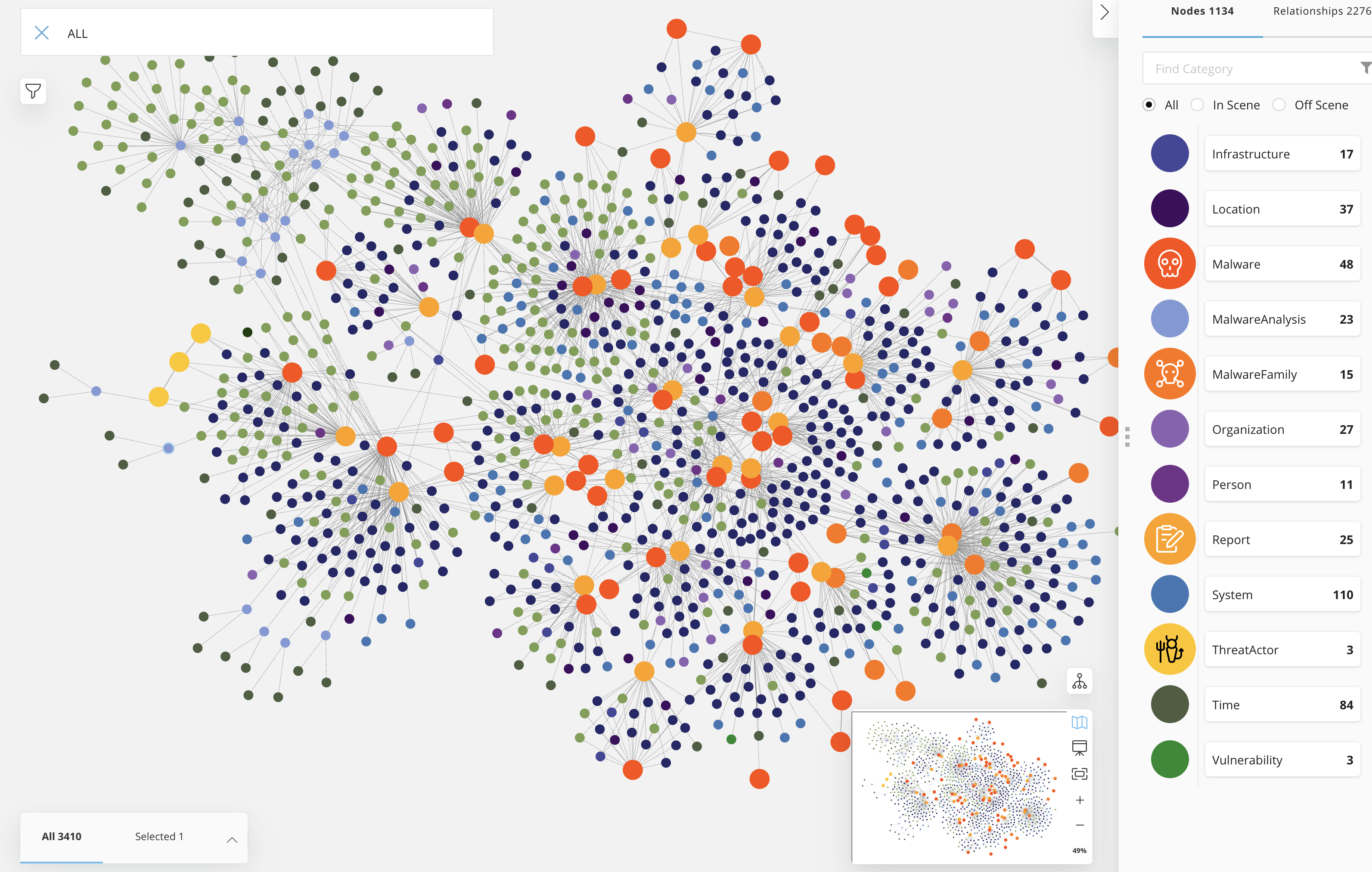}}
   \caption{MalKG}
    \label{fig:AllTriples}
\end{figure}

\begin{figure}[hbt!]
    \centering
   \frame{
   \includegraphics[width=.4\textwidth,height=.25\textwidth]{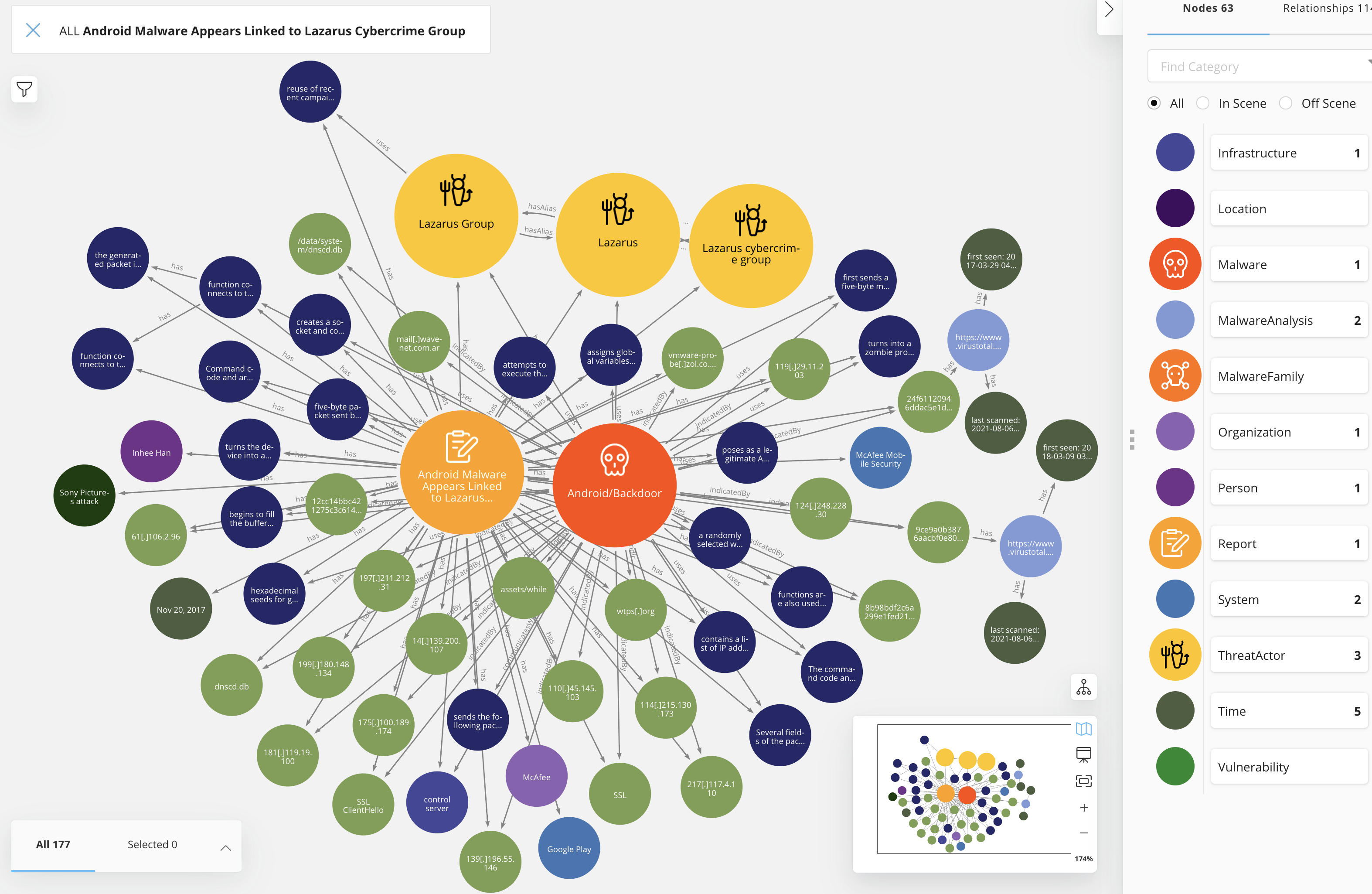}
  }    
   \caption{Complete sub-graph from a single CTI report}
    \label{fig:ReportTriples}
\end{figure}

\begin{figure}[hbt!]
    \centering
 \frame{
   \includegraphics[width=.22\textwidth,height=.17\textwidth]{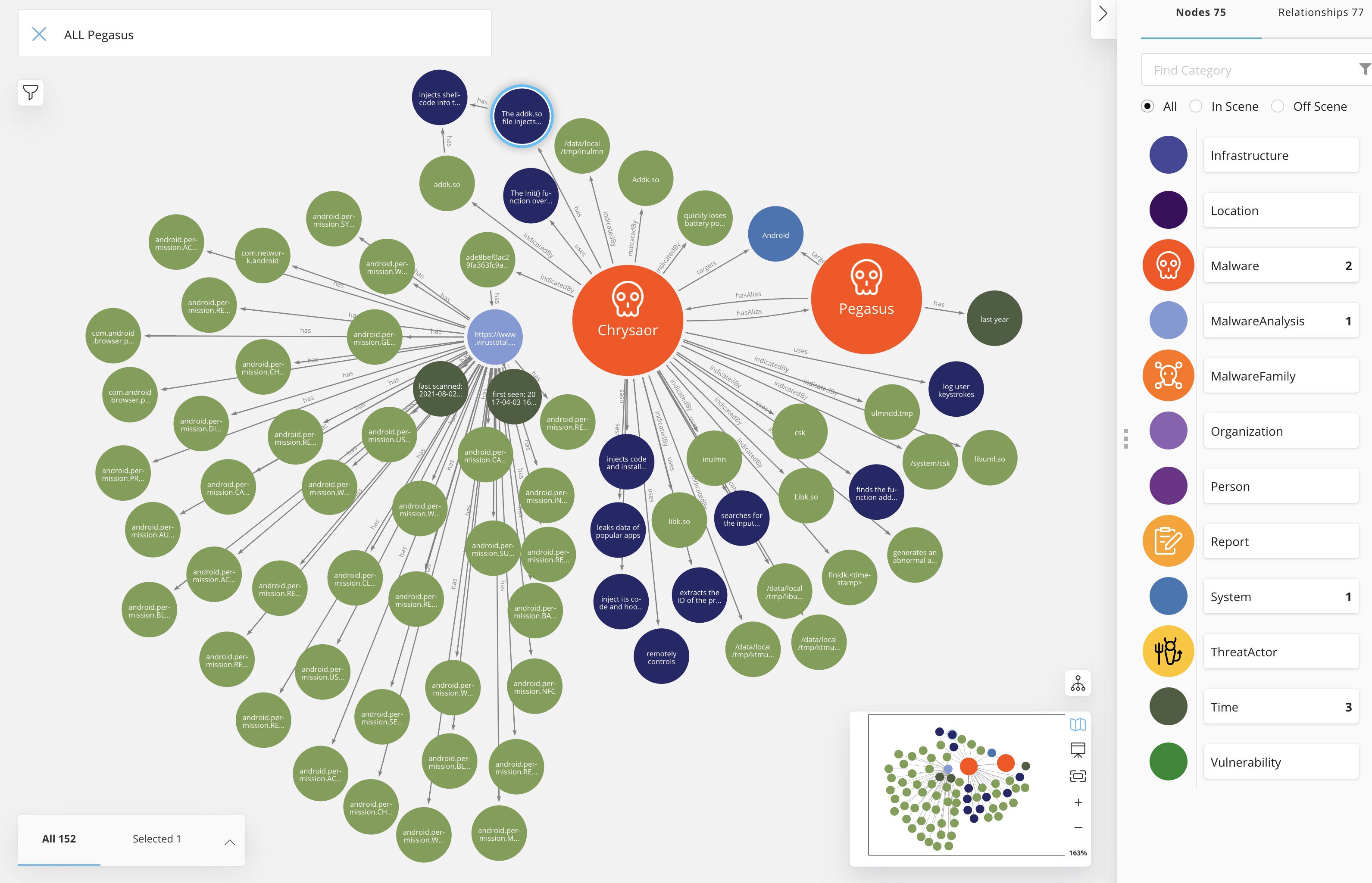}}
  \hspace{0.2cm}
  \frame{
   \includegraphics[width=.22\textwidth,height=.17\textwidth]{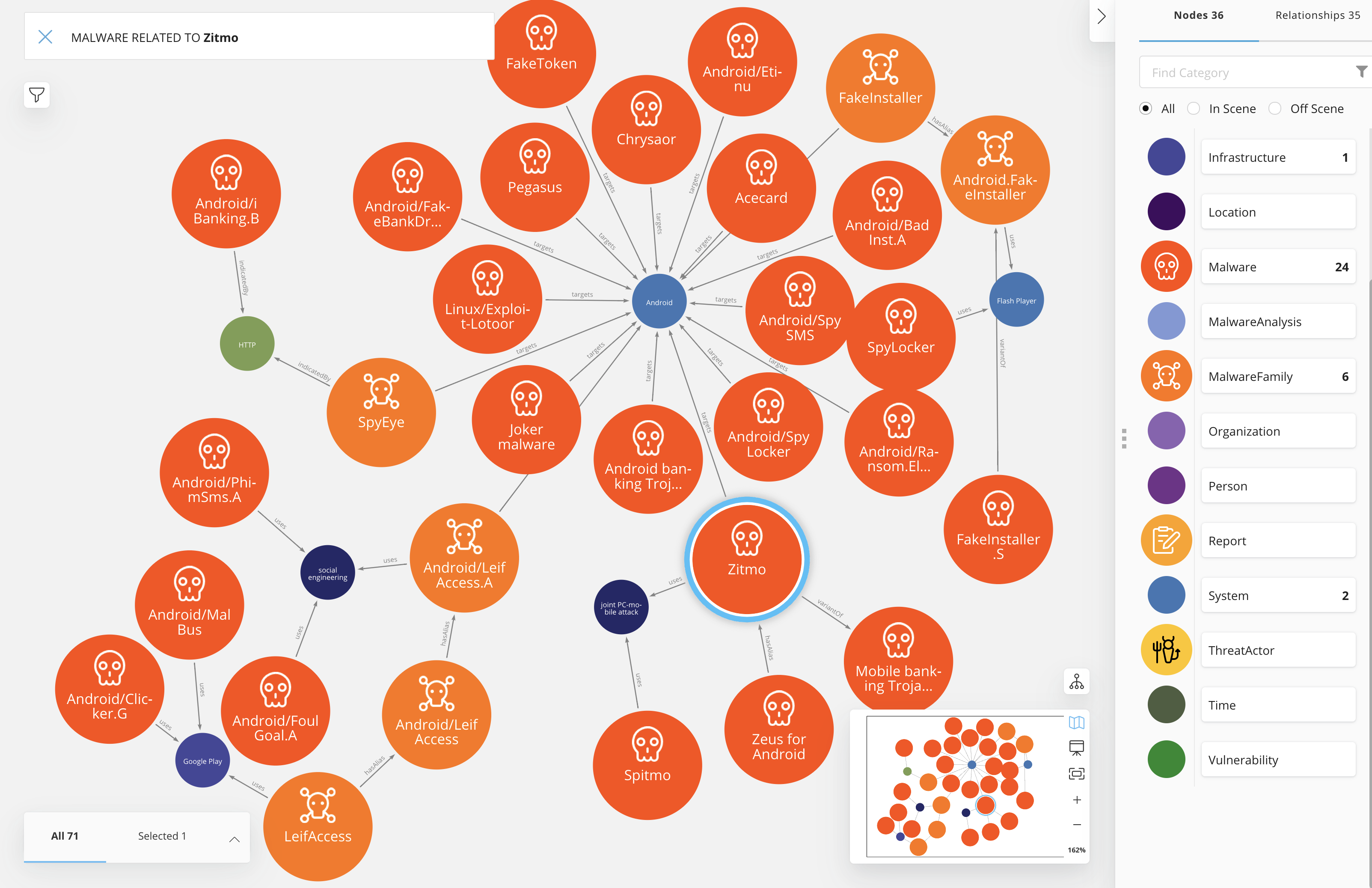}
  }   

   \caption{Subgraph of a single malware (left), path from Zitmo malware to other related malware (right).}
    \label{fig:snap_kg_2}
\end{figure}

Our demonstration describes the ontology MalONT2.0 and briefly compares it with other ontologies, namely UCO \cite{uco}, MalONT (our prior work)\cite{rastogi2020malont}, and Swimmer ontology\cite{swimmer}-- arguably the first malware ontology. These ontologies have been chosen for comparison based on the competency questions defined in Section \ref{newMalONT}. We also have a live demonstration of MalONT2.0 ontology prepared, as well as queries for the knowledge graph. A few queries of the knowledge graph are visualized using Neo4j Bloom\footnote{https://neo4j.com/developer/neo4j-bloom/} in figures \ref{fig:AllTriples}, \ref{fig:ReportTriples}, and \ref{fig:snap_kg_2}.

\subsection{MalONT2.0 based annotations}
\begin{figure}[hbt!]
 \frame{
   \includegraphics[width=.22\textwidth,height=.1\textwidth]{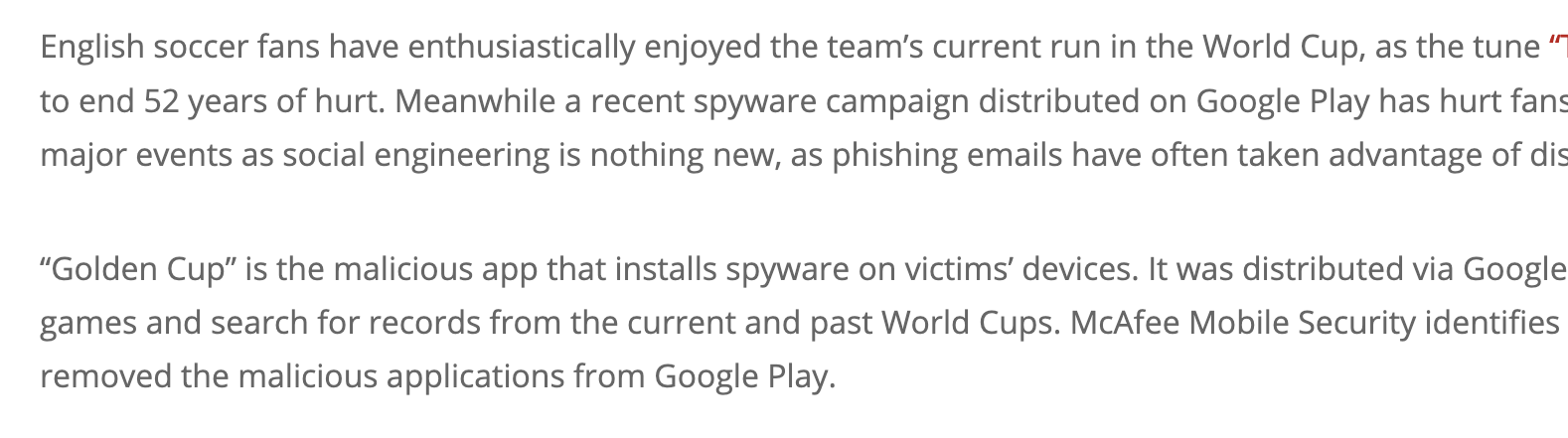}}
   \hspace{0.15cm}
   \frame{
      \includegraphics[width=.22\textwidth,height=.1\textwidth]{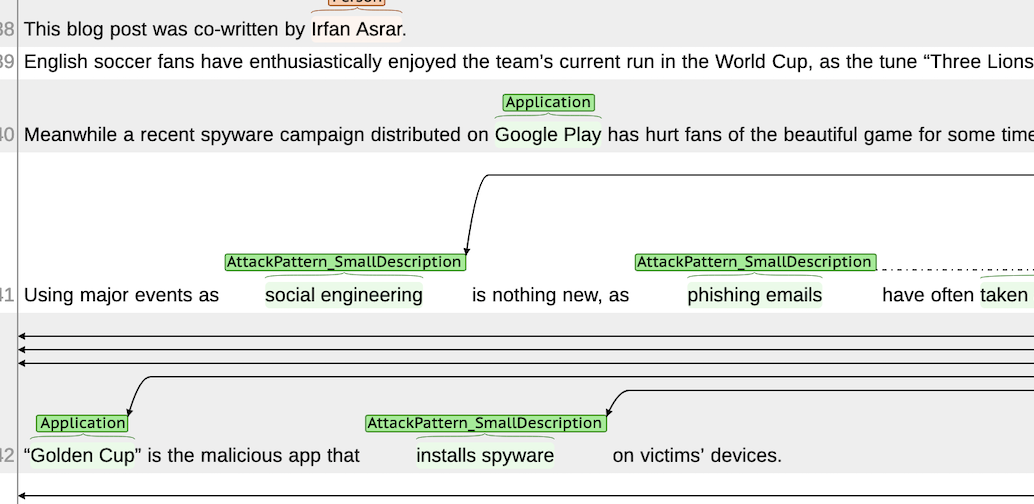}}   
      
      \caption{McAfee Blog snippet (left), BRAT's interface (right)}
    \label{fig:threatReport}
\end{figure}

First we will demonstrate (a) MalONT2.0 in protege showing all the classes, sub-classes, and their descriptions, (b) configuration of BRAT\footnote{https://brat.nlplab.org/} using the OWL file generated by MalONT2.0 prior to annotating CTI reports, (c) semantic features in MalONT2.0 based on STIX2.1 framework and the syntactic features and factual data based on VirusTotal, (d) triple generation in the form of RDF from the annotated text, and assignment of classes and relations. See Figure~\ref{fig:threatReport} for a snippet of a McAfee threat report on a spyware, "Golden Cup".

\subsection{Demonstrate MalKG}
We will demonstrate (a) MalKG generated using only annotated triples, (b) MalKG generated using all CTI reports collected so far using the dynamically growing platform, (c) an entire sub-graph of a malware extracted from the larger KG. This will include all the triples connected to this malware including those generated from the CTI reports and VirusTotal (see Figure \ref{fig:snap_kg_2}).

\subsection{Run queries on MalKG}
We will demonstrate live queries on the MalKG through the Neo4j platform. Queries will include searching for entities such as a malware, an application, or a vulnerability and demonstrate the kinds of sub-graphs that can be drawn from them (see Figure \ref{fig:ReportTriples} and \ref{fig:snap_kg_2}).

\section{Conclusion and Future Work}
We present MalONT2.0 ontology for capturing contextual threat intelligence from heterogeneous sources. CTI reports provide semantic information which is used to instantiate classes and relations of MalONT2.0. VirusTotal provides additional syntactic information for hashes that occur in the CTI reports, and the graph generated from this is appended to the existing MalKG. A dynamic knowledge graph, TINKER, is also proposed. For future work, we plan to perform validations on triples generated for TINKER. We also plan to demonstrate the use on TINKER for forecasting threat vectors. 

\section{Acknowledgement}
This work is supported by the IBM AI Research Collaboration (AIRC). The authors would like to thank RPI researchers Erin Turnbull and Yueting Liao for their support.
\bibliographystyle{ACM-Reference-Format}
\bibliography{sample-base}


\begin{thebibliography}{7}


\ifx \showCODEN    \undefined \def \showCODEN     #1{\unskip}     \fi
\ifx \showDOI      \undefined \def \showDOI       #1{#1}\fi
\ifx \showISBNx    \undefined \def \showISBNx     #1{\unskip}     \fi
\ifx \showISBNxiii \undefined \def \showISBNxiii  #1{\unskip}     \fi
\ifx \showISSN     \undefined \def \showISSN      #1{\unskip}     \fi
\ifx \showLCCN     \undefined \def \showLCCN      #1{\unskip}     \fi
\ifx \shownote     \undefined \def \shownote      #1{#1}          \fi
\ifx \showarticletitle \undefined \def \showarticletitle #1{#1}   \fi
\ifx \showURL      \undefined \def \showURL       {\relax}        \fi
\providecommand\bibfield[2]{#2}
\providecommand\bibinfo[2]{#2}
\providecommand\natexlab[1]{#1}
\providecommand\showeprint[2][]{arXiv:#2}

\bibitem[\protect\citeauthoryear{Barnum}{Barnum}{2012}]%
        {stix}
\bibfield{author}{\bibinfo{person}{Sean Barnum}.}
  \bibinfo{year}{2012}\natexlab{}.
\newblock \showarticletitle{Standardizing cyber threat intelligence information
  with the Structured Threat Information eXpression (STIX)}.
\newblock \bibinfo{journal}{\emph{Mitre Corporation}}  \bibinfo{volume}{11}
  (\bibinfo{year}{2012}), \bibinfo{pages}{1--22}.
\newblock


\bibitem[\protect\citeauthoryear{Connolly, Davidson, and Schmidt}{Connolly
  et~al\mbox{.}}{2014}]%
        {connolly2014trusted}
\bibfield{author}{\bibinfo{person}{Julie Connolly}, \bibinfo{person}{Mark
  Davidson}, {and} \bibinfo{person}{Charles Schmidt}.}
  \bibinfo{year}{2014}\natexlab{}.
\newblock \showarticletitle{The trusted automated exchange of indicator
  information (taxii)}.
\newblock \bibinfo{journal}{\emph{The MITRE Corporation}}
  (\bibinfo{year}{2014}), \bibinfo{pages}{1--20}.
\newblock


\bibitem[\protect\citeauthoryear{Pujara, Miao, Getoor, and Cohen}{Pujara
  et~al\mbox{.}}{2013}]%
        {pujara2013knowledge}
\bibfield{author}{\bibinfo{person}{Jay Pujara}, \bibinfo{person}{Hui Miao},
  \bibinfo{person}{Lise Getoor}, {and} \bibinfo{person}{William Cohen}.}
  \bibinfo{year}{2013}\natexlab{}.
\newblock \showarticletitle{Knowledge graph identification}. In
  \bibinfo{booktitle}{\emph{International Semantic Web Conference}}. Springer,
  \bibinfo{pages}{542--557}.
\newblock


\bibitem[\protect\citeauthoryear{Pustejovsky and Stubbs}{Pustejovsky and
  Stubbs}{2013}]%
        {pustejovsky2013natural}
\bibfield{author}{\bibinfo{person}{James Pustejovsky} {and}
  \bibinfo{person}{Amber Stubbs}.} \bibinfo{year}{2013}\natexlab{}.
\newblock \bibinfo{title}{Natural Language Annotation for Machine Learning.
  OReilly Media}.
\newblock
\newblock


\bibitem[\protect\citeauthoryear{Rastogi, Dutta, Zaki, Gittens, and
  Aggarwal}{Rastogi et~al\mbox{.}}{2020}]%
        {rastogi2020malont}
\bibfield{author}{\bibinfo{person}{Nidhi Rastogi}, \bibinfo{person}{Sharmishtha
  Dutta}, \bibinfo{person}{Mohammed~J Zaki}, \bibinfo{person}{Alex Gittens},
  {and} \bibinfo{person}{Charu Aggarwal}.} \bibinfo{year}{2020}\natexlab{}.
\newblock \showarticletitle{MalONT: An ontology for malware threat
  intelligence}. In \bibinfo{booktitle}{\emph{International Workshop on
  Deployable Machine Learning for Security Defense}}. Springer,
  \bibinfo{pages}{28--44}.
\newblock


\bibitem[\protect\citeauthoryear{Swimmer}{Swimmer}{2008}]%
        {swimmer}
\bibfield{author}{\bibinfo{person}{Morton Swimmer}.}
  \bibinfo{year}{2008}\natexlab{}.
\newblock \showarticletitle{Towards an ontology of malware classes}.
\newblock \bibinfo{journal}{\emph{Online] January}}  \bibinfo{volume}{27}
  (\bibinfo{year}{2008}).
\newblock


\bibitem[\protect\citeauthoryear{Syed, Padia, Finin, Mathews, and Joshi}{Syed
  et~al\mbox{.}}{2016}]%
        {uco}
\bibfield{author}{\bibinfo{person}{Zareen Syed}, \bibinfo{person}{Ankur Padia},
  \bibinfo{person}{Tim Finin}, \bibinfo{person}{Lisa Mathews}, {and}
  \bibinfo{person}{Anupam Joshi}.} \bibinfo{year}{2016}\natexlab{}.
\newblock \showarticletitle{UCO: A unified cybersecurity ontology}. In
  \bibinfo{booktitle}{\emph{Workshops at the Thirtieth AAAI Conference on
  Artificial Intelligence}}.
\newblock


\end{thebibliography}

\appendix









\end{document}